\documentstyle[12pt,aasms4]{article}
\def\bib{\par\noindent\hangindent=3mm\hangafter=1}
\def\mv{M_V}

\def\msol{M_\odot}
\def\lsol{L_\odot}
\def\mbol{M_{\rm bol}}
\def\mvol{{\rm M}_\odot{\rm pc}^{-3}}

\def\te{T_{\rm eff}}

\def\kms{\rm km.s^{-1}}
\def\simgr{\,\hbox{\hbox{$ > $}\kern -0.8em \lower 1.0ex\hbox{$\sim$}}\,}
\def\simle{\,\hbox{\hbox{$ < $}\kern -0.8em \lower 1.0ex\hbox{$\sim$}}\,}

\begin{document}

\title{Is Galactic Dark Matter White ?}
\author{ G. Chabrier}
\affil{Centre de Recherche Astrophysique de Lyon (UMR CNRS 5574),\\
 Ecole Normale Sup\'erieure de Lyon, 69364 Lyon
Cedex 07, France\\
(chabrier@ens-lyon.fr)}

\begin{abstract}
We calculate the expected white dwarf luminosity functions and
discovery functions
in photometric passbands, if these stellar remnants provide
a substantial fraction of the sought Galactic dark matter, as
suggested on various observational grounds. We demonstrate the extremely rapid
variation 
of the photometic signature of halo white dwarfs with time and thus the powerful diagnostic of
white dwarf colors to determine the age of the Galactic halo.
We also consider the various indirect constraints implied by a white dwarf dominated
halo. These calculations
will guide present and future
observational projects at faint magnitudes. This
will enable
us to determine not only the nature of the Galactic dark matter
but also the age and the initial conditions of the Galaxy formation.
\end{abstract}

\section{Introduction}

The nature of Galactic dark matter remains a major puzzle. 
The MACHO collaboration has now more than 12 events observed towards the LMC
(Alcock et al., 1998).
They confirm an optical depth $\tau =2.9^{+1.4}_{-0.9}\times 10^{-7}$ and
a mass of dark objects within 50 kpc M$_{50}=2.0^{+1.2}_{-0.7}\times
10^{11}\,
\msol$, {\it at least} 20\% of the mass of the Galaxy within this limit.
The time
distribution of the microlensing
events is narrow ranged  ($\hat t\sim$ 20-70 days) with
$\langle t_e \rangle\approx 40$ days (for the Einstein radius crossing time). Assuming a standard $1/r^2$ isothermal
halo, this means that 
a fraction $\sim 60\pm 20 \%$ of the sought
Galactic dark mass might be in the form of objects with
an average mass
$\langle  m \rangle = 0.5^{+0.3}_{-0.2}\,\msol$.

Brown dwarfs and low-mass stars ($m\simle 1\,\msol$) are
excluded as a significant halo population
both by the microlensing time distribution and by the star
count analysis of the Hubble Deep Field (HDF) at faint magnitudes, yielding
for the dark halo a maximum stellar plus brown dwarf density $\rho_{h_{\star+BD}} \simle 0.001\times \rho_{h}$, where $\rho_h\approx 10^{-2}\,\mvol$ is the
dark
halo local dynamical density
(Chabrier and M\'era,
1997; Gould, Flynn and Bahcall, 1998).
It is important to stress that both these
observational constraints on the stellar content
of the Galactic dark halo
and the narrow range of time distribution of the events observed
towards the LMC imply 
an initial mass function (IMF) {\it different} from a Salpeter
one below $\sim 1\,\msol$.
The recent determination of the Galactic local volume density (Cr\'ez\'e et al.,
1998) leaves essentially no
room for dark matter in the disk, and thus strongly favors a
non-dissipative component for the Galactic dark matter. 
The suggestion that the events
might be due to an intervening
stellar population (dwarf galaxy or tidal debris)
along the
light-of-sight (Zaritsky and Lin, 1997; Zhao, 1998) is still controversial
(Beaulieu and Sackett, 1998;
Alcock et al., 1997).
Under these circumstances, WD's remain the most favorable candidates for
the observed microlensing events and the Galactic baryonic dark matter, in
particular after the recent
demonstration that some of the faint
blue objects in the HDF are consistent with very cool ($\te \simle 4000$ K)
H-rich atmosphere WD's (Hansen, 1998).
In this paper we calculate the expected halo white dwarf luminosity
functions (WDLF) and discovery functions for different halo ages,
in various photometric passbands, for a halo WD population
consistent with the dark mass inferred from the MACHO observations.
This works extends beyond our initial work (Chabrier et al., 1996)
by including new atmosphere models and spectral colors appropriate for the halo population.

\section{Halo white dwarf luminosity function}

For WD's to provide a mass fraction $X_{WD}$ of the halo local dynamical density
$\rho_h$, the WDLF must be
normalized as :

\begin{equation}
\int n\,dM_{bol}={X_{WD}\over \langle m_{WD} \rangle}\times \rho_h
\end{equation}

\noindent The WDLF reads (Chabrier et al., 1996; Adams \& Laughlin, 1996):

\begin{equation}
{dn\over d\mbol}=\int_{m_{inf(L)}}^{m_{sup(L)}} \psi [t(m,L)]\times
\phi(m)\times ({\partial t_{cool}(m,L)\over \partial \mbol}) dm
=({\partial t_{cool}\over \partial \mbol}) \,{dm \over dt}\, \nu(t)
\end{equation}

\noindent where $m$ is the WD progenitor mass and $\psi(t)$ and $\phi(m)$ denote the stellar formation rate (SFR) and
the IMF.
The second equality stems from the assumption
that the star formation burst at the early epoch of the Galaxy is
much shorter than the age of the halo, so that the SFR can be approximated
by a Dirac function $\psi(t)=\delta(t-t_0)$. In that case
$\nu(t)$ is simply
the number of stars such that $t_{MS}+t_{cool}=t_h$, where $t_h$ is the
halo age
and $t=t_{MS}+t_{cool}$ is the WD total age, i.e. its cooling
time plus the main sequence lifetime of its progenitor.

The IMF in (2) must fulfill several constraints: the HDF observations 
imply $\rho_{(m<1.0)} << 0.01\times \rho_h$ (see above) and the presence of
type II supernovae at finite redshift (Miralda-Escud\'e and Rees, 1997)
imply a finite fraction of stars above $\sim 8\,\msol$.
We elected a truncated power-law function (Larson, 1986; Chabrier et al., 1996):

\begin{equation}
\phi(m)={dn\over dm}=A\, exp^{ - ({\overline {m}\over  m})^{\beta} }\, m^{-\alpha}
\end{equation}

Equation (3) approaches a power-law form $m^{-\alpha}$ at large
masses and can easily be ajusted to reproduce any observed SNII rate, while the
integral of the mass function,
which is what matters in the present context, is determined essentially by
the peak $m_p=(\beta/ \alpha)^{1/\beta}\, \bar m$ and by the parameter $\beta$.
In order to examine the dependence of the results upon the IMF parameters,
the present calculations have been conducted with
$\bar m=3.5\,\msol$, $\beta=3.0$, $\alpha=5.0$, hereafter IMF1, which yields an average WD mass $\langle m_{WD} \rangle\approx 0.8 \,\msol$
and
$\bar m=2.4\,\msol$, $\beta=3.0$, $\alpha=5.0$, hereafter IMF2, which yields $\langle m_{WD}\rangle \approx 0.7 \,\msol$.
The slightly larger average mass for halo WD's than for disk WD's is motivated
by (i) the smaller mass-loss during evolution for metal-poor stars (Maeder, 1992) and (ii) by the fact that the faintest
observed disk WD's have masses $\sim 0.7$-$0.8\,\msol$ (Leggett, Ruiz and Bergeron, 1998). Both IMF1 and IMF2 yield a mass-to-light ratio $M/L>>100$,
as required for a dominantly baryonic halo. Note in passing that this type of IMF provides a natural
explanation for the lack of zero-like metallicity stars in the Galaxy, the so-called G-dwarf problem.

The present calculations include the most recent atmosphere profiles and
synthetic spectra calculations for pure hydrogen atmosphere (so-called DA) WD's (Saumon and Jacobson, 1998),
and the
most updated WD interior physics, C/O profiles (Salaris et al., 1997),
equation of state and 
crystallization along evolution
(Segretain et al., 1994; Chabrier et al., 1996).
To illustrate the rapid cooling of halo WD's, we found out that although
for $t_h\simle$12 Gyr
the entire WD population is brighter than $\mv=20$, i.e. $\mbol \approx 21$-22,
after 14, 15 and 16 Gyr,
only $\sim$ 80\%, 60\% and 25\%, respectively, of the total WD population remains brighter than this magnitude. The more massive WD's have cooled fainter.
For pure-He (more transparent) atmosphere WD's, the situation is even more dramatic and after $\sim 8 $ Gyr, the majority of
these stars have cooled fainter than $\mbol=21$ and will thus escape detection.
However,
using the Alcock and Illarionov (1980) accretion formula,
$dm/dt \approx 10^{-20}\,({m\over 0.5\,\msol})\,\msol$.yr$^{-1}$, these stars will accrete
$\sim 10^{-13}\,\msol$ of hydrogen, i.e. about a photosphere mass, during each disk crossing and will thus very likely cool like H-rich atmosphere WD's.

Figure 1 displays the expected halo WDLF for DA's with the IMF1 (solid line), normalized to
$X_{WD}=50\%$, in the most favored V-band, for $t_h=14, 15$ and
16 Gyr, about the age of the oldest globular clusters (Pont et al., 1998)\footnote{Note that t$\approx$15 Gyr corresponds to the age of the universe for $\Omega_{matter}$=0.24, $\Omega_\Lambda$=0.62 and $H_0$=70 km s$^{-1}$ Mpc$^{-1}$, the presently most favored parameters.}.
The end of the observed disk WDLF is also
displayed, as well as the observed WD's with halo-like kinematics, i.e.
$v_{tan}\ge 250\,\kms$ (Liebert, Dahn and Monet, 1988). 
It is important to mention the sensitivity of the
calculations upon the different input parameters, namely
(i) the progenitor mass - WD mass relationship, (ii) the progenitor
main sequence lifetime and (iii) the IMF.
All the present calculations were done with the disk
characteristic relationships (Iben and Tutukov, 1984). Although modifications of
the two first relationships were found to affect only moderately the
WDLF, the IMF is determinant.
This is illustrated by the dashed line in Fig. 1, which displays
the results of the same calculations with the IMF2.
Although
the peak of the WDLF is not affected significantly
(shifted by $\sim$0.5-1 mag), the bright part of the WDLF, which
stems from the low-mass tail of the IMF,
has changed by orders of magnitude.
This shows the extreme sensitivity of the bright part of the halo WDLF to the
ill-constrained low-mass end of the IMF and thus the necessity to
observed the bulk of the halo WDLF in order to constrain the IMF of the
Galactic halo.

Figure 2 shows WD evolution in a color-magnitude diagram (CMD) and
illustrates the rapid variation of halo WD optical colors with time. The photometric observation of identified halo WD's will thus provide a powerful
diagnostic to determine the age of the Galactic halo. It also provides
complementary information to the observation of the WDLF.
As shown in Fig. 1, a very small number of objects at $\mv \simle 20$ might reflect
either a negligible fraction of halo WD's or a halo age older than 16 Gyr.
The CMD will resolve this ambiguity by providing a determination of the
age of the objects.

\section{Indirect constraints}

A Galatic halo composed dominantly of stellar remnants like WD's implies
several constraints on the Galaxy genesis and evolution.
The mass fraction of returned gas between the progenitor and the WD with respect to the remnant mass can be written
(still assuming a Dirac-function for the SFR):

\begin{equation}
R(t)={ \int_{\sim 0.5\msol}^{\sim 8 \msol} \{ m-m_{WD}(m) \} \phi(m) dm \over
\int_{\sim 0.5\msol}^{\sim 8 \msol} m_{WD} \phi(m) dm }
\end{equation}

The IMF1 and IMF2 yield $R(t)=3.5$ and 2.7, respectively. The total mass of returned gas is thus
$M_{gas}\approx 1.5\times 10^{12}\,\msol$ if $X_{WD}=0.5$.
Most of this gas will appear in the
Lyman-$\alpha$ forest and the
total mass of baryons in the leftover gas and in the stellar remnants 
is bound by the total baryon density, {\it for the same redshift value} :
$\Omega_{rem}(z) + \Omega_{Ly\alpha}(z)\le \Omega_B\approx (0.019\pm 0.002)h^{-2}$ (Fields et al., 1998).
Present determinations yield $\Omega_{Ly_\alpha}\approx 0.02$ at $z\sim 2$
(Petitjean et al., 1993).
The leftover gas is likely to be ejected in the intergalactic medium (IGM) by a wind
due to the primordial generation of SNII or by the more efficient merging mechanism (Gnedin, 1998).
This implies the presence of
some hot ($\sim$ 10$^5$-$10^6$ K) X-ray gas in the Local Group (LG). Although there are only hints for the
presence of such gas in the LG,
its presence has been established in other galaxy groups and in the intracluster medium
(see e.g. Mushotzky et al., 1996).
The total mass of metals ejected during the envelope ejection phase of the WD
progenitors is :
\begin{equation}
M_Z=y_Z\times \int m\,\phi(m)\,dm
\end{equation}

\noindent where $y_Z$ denotes the metal stellar yields.
As noted by Gibson and Mould
(1997), a strongly peaked IMF around 2 $\msol$ will
produce $[C,N/O]$
abundances during the AGB phase about 10 times the
observed value in halo stars.
This result, however, depends entirely on the assumed
stellar yields $y_Z$. Gibson and Mould
used yields appropriate for $Z\simgr 10^{-2}\times Z_\odot$.
Stellar evolution
calculations for zero-like metallicities (Chieffi and Tornamb\'e, 1984;
Fujimoto et al., 1984) show that for a central degenerate core $M\simgr
0.77\,\msol$, thermal pulses along the asymptotic giant branch do not occur,
so that the bottom of the convective envelope never reaches the carbon-enriched region
and remains unpolluted. This leads to no C-enhancement of the interstellar medium during the
planetary nebulae phase.
For zero-like metallicities, this core mass corresponds to $m \simgr 3\,\msol$,
a condition statisfied for most of the stars with IMF1 and marginally
satisfied by the IMF2.
The C-enrichment constraint might thus be relaxed for primordial
stars, depending on the IMF.
On the other hand,
the presence of an unexpected level of heavy-element enrichment in the IGM at high
redshift has been established observationaly (Cowie and Songaila, 1998).
The identification of AGB stars as the origin of some of these elements
would bring immediate support to the halo WD scenario.

The halo WD scenario seemed to have been excluded on
the basis of the observed rate of type Ia SN in galaxies (Smecker \& Wyse, 1991). These
calculations, however,
assumed that the merging of two WD's whose total mass exceeds the
Chandrasekhar mass
would produce a SNIa.
Recent calculations (Segretain, Chabrier and
Mochkovitch, 1997) seem to exclude, or at least
strongly unfavor the formation of SNI by this scenario.
These calculations seem to be supported on various observational grounds,
suggesting that the rate of SNIa from merging
WD's has been significantly overestimated (see Segretain et al., 1997;
Maxted and Marsh, 1998). Note also that
usual assumptions about binary parameters in the Galactic disk (mass loss, orbital radius, rate) are likely to be
irrelevant under the completely different primordial halo conditions.
Charlot and Silk (1995) showed that a halo WD population
such that $X_{WD}\simgr 10\%$ would correspond to a progenitor light at
redshift $z\le 3.5$ at odd with the observational constraints.
These
calculations, however, were done for a halo age $t_h=$13 Gyr and for stellar evolution
models with solar metallicity. Low metallicity stars are brighter for the
same mass and thus evolve more quickly. This - and an older halo age - weakens the Charlot-Silk
constraint or at least pushes it to larger redshifts.
The background light of the progenitors of a WD-dominated halo is constrained also by
the total amount of energy
distribution in the universe determined by the DIRBE
observations (Guiderdoni et al., 1997).
The IMF1 (resp. IMF2) yields
$\langle m\rangle\sim 4\,\msol$ (resp. $\sim 3\,\msol$) for the progenitors, i.e. $\langle L\rangle\sim 80\times \lsol$
(resp. $\sim 30\times \lsol$) over $\sim$1 Gyr.
Since the mass of the LG is $M_{LG}\sim 2\times 10^{12}\,\msol$, this
yields $\langle L_{LG}\rangle\simle 10^{47}$ erg/s.
The radius of the LG protogalactic region
was $R\sim 1$ Mpc (Peebles, 1993), which yields a surface brightness at redshift $z$,
$\mu_z\simle (2.6\times 10^{-4})/(1+z)^4$ erg s$^{-1}$ cm$^{-2}$ sr$^{-1}$.
For a halo formed at redshift $z\simgr 4$, this yields a peak
distribution $\mu_z \simle 4.0\times 10^{-7}$
erg s$^{-1}$ cm$^{-2}$ sr$^{-1}$ at $\lambda_z= \lambda_0(1+z)\simgr
1\,\mu$m, about a factor 50 below the DIRBE limit.

\section{Predicted counts}

The number of WD's of absolute magnitude $M_\lambda$ per arcmin$^2$ observable in a field of longitude and
latitude $(l,b)$,
for a limit magnitude $m_l$ reads:

\begin{equation}
N_{WD}={1\over 3600} ({\pi \over 180})^2 R_0^2\times \{ \int^{M_\lambda} ndM
\,\int_0^{d_{max}(M_\lambda)} {r^2\, dr\over R_0^2+r^2-2rR_0 \cos b \, \cos l } \}
\end{equation}

\noindent where $R_0=8.5$ kpc is the galactocentric position of the Sun
and $\log d_{max}(M_\lambda) \, [pc]=0.2(m_l-M_\lambda)+1$.
Using the WDLF's determined in the present calculations for pure DA WD's and X$_{WD}$=50\%, at most $N\sim 2$ WD's should have been
expected in the HDF (4.4 arcmin$^2$) at V$\approx$I$\approx$28 for $t_h=14$ Gyr, whereas less than 1 is expected with the FORS1 VLT telescope
($V=26$; 6.8 arcmin$^2$).
In the {\it total}
available field of the french EROSII survey (250$^{o^2}$), about $N\sim 26$, 2 and 0.01 WD's are expected for
halo ages $t_h=14, 15$ and 16 Gyr, respectively at $I_{lim}= 20$ in the appropriate (age-dependent) colors
(see Fig. 2). These numbers are multiplied by
a factor $\sim 4$ for $I_{lim}=21$, a factor $\sim 15$ for $I_{lim}=22$, and a factor $\sim 1000$
for $I_{lim}$=25.
Note that for $t_h \ge 16$ Gyr, the expected number of WD's remains essentially zero or a few.
A useful tool for observers
is the so-called discovery function $D(M_\lambda)$, i.e. the number of WD's observable over the
whole sky. 
For a survey limited to nearby halo WD's, the density can
be considered as constant, and
$D(M_\lambda)= {4\pi \over 3} d_{max}^3(M_\lambda) n(M_\lambda) $.
Figure 3 shows the expected discovery functions in the V-band
for $X_{WD}=0.5$ and different halo ages and magnitude limits.

\section{Conclusion}

We have shown in this paper that most of identified baryonic components
are
unlikely to provide a substantial contribution to the Galactic missing
mass, except if they are distributed inhomogeneously, which implies a
varying mass-to-light ratio in the Galactic dark halo. White dwarfs,
although raising important issues about the Galaxy formation and
evolution, remain the most plausible
candidates
to explain the observed microlensing events and might provide a substantial
fraction of the sought baryonic dark matter. 
The luminosity functions, discovery functions and star counts calculated
in the present paper will
guide ongoing and future
observational projects at faint magnitudes.
As shown in Fig. 2, the photometric
observation of halo WD's will provide a powerful diagnostic to determine
the age of the Galactic halo.
If WD's do indeed account for a large fraction of the Galactic missing mass, they solve the dilemna
of the missing baryons at $z=0$. In that case the baryonic missing mass is composed essentially
of these stellar remnants in galactic halos and
of the intergalactic left-over gas from the progenitors.

{\bf Ackowledgments} : I am deeply endebted to D. Saumon and G. Fontaine for allowing me
to use results prior to publication, and to
P. Petitjean and U. Fritze for very useful conversations. My gratitude to M. Hernanz for sending the C/O profiles. I am also grateful to the ESO, where most
of this work was accomplished, for a visiting
scientist position.

\vfill\eject
\section*{References}

\bib Adams, F.C., and Laughlin, G., 1996, \apj, 468, 586
\bib Alcock, C. et al., 1998, in {\it The Galactic Halo}, ASP Conf. Ser., 666, 1999
\bib Alcock, C., et al., 1997, \apjlett, 490, L59
\bib Alcock, C., and Illarionov, 1980, \apj, 235, 541
\bib Beaulieu, J.P., and Sackett, P., 1998, \aj, 116, 209
\bib Chabrier, G., Segretain, L. \& M\'era, D., 1996, \apj, 468, L21
\bib Chabrier, G., and M\'era, D., 1997, \aap, 328, 83
\bib Charlot, S., and Silk, J., 1995, \apj, 445, 124
\bib Chieffi and Tornamb\'e, 1984, \apj, 287, 745
\bib Cowie, L.L., and Songaila, A., 1998, Nature, 394, 44
\bib Cr\'ez\'e, M., Chereul, E., Bienaym\'e, O., and Pichon, C.,
1998, \aap 329, 920
\bib Fields, B., Freese, K., and Graff, D., 1998, New Astronomy, 3, 347
\bib Fujimoto, M.Y., Iben, I., Chieffi, A., and Tornamb\'e, A., 1984, \apj, 287, 749
\bib Gibson, B., and Mould, J., 1997, \apj, 482, 98
\bib Gnedin, N.Y., 1998, \mnras, 294, 407
\bib Gould, A., Flynn, C., and Bahcall, J., 1998, \apj, 503, 798
\bib Guiderdoni, B., Bouchet, F., Puget, J-L, Lagache, G., and
Hivon E., 1997, Nature, 390, 257
\bib Hansen, B.M.S., 1998, Nature, 394, 860
\bib Iben, I., and Tutukov, A., 1984, 282, 615
\bib Larson, R. B., 1986, \mnras, 218, 409
\bib Leggett, S., Ruiz, M.T., and Bergeron, 1998, \apj, 497, L294
\bib Liebert, J., Dahn, C.C.\& Monet, D.G., 1988, \apj, 332, 891
\bib Maeder, A., 1992, \aap, 264, 105
\bib Maxted, P., and Marsh, T, 1998, \mnras, 296, L34
\bib Miralda-Escud\'e, J., \& Rees, M.,J., 1997, \apj, 478, L57
\bib Mushotzky, R.F., et al., 1996, \apj, 466, 686
\bib Peebles, 1993, {\it Principle of Physical Cosmology}, U. of Princeton Press
\bib Petitjean, P., Webb, J.K., Rauch, M., Carswell, R.F., and
Lanzetta, K., 1993, \mnras, 262, 499
\bib Pont, F., Mayor, M., Turon, C., and Van den Berg, D., 1998, \aap, 329, 87
\bib Salaris, M., Dom\'inguez, I., Garc\'ia-Berro, E., Hernanz, M., and
Mochkovitch, R., 1997, \apj, 486, 413
\bib Saumon, D., and Jacobson, S.B., 1998, \apj, in press
\bib Segretain, L. Chabrier, G.,
and Mochkovitch, R., 1997, \apj, 481, 355
\bib Segretain, L., Chabrier, G., Hernanz, M., Garc\' ia-Berro,
E., Isern,
J. \& Mochkovitch,
R., 1994, \apj, 434, 641
\bib Smecker, T.A., and Wyse, R., 1991, \apj, 372, 448
 C.M., Silk, J., Wood, M.A., and Winget, D.E., 1990, \apj, 358, 164
\bib Zaritsky, D., and Lin, D.N.C., 1997, \aj, 114, 254
\bib Zhao, H., 1998, \apjlett 500, L149; 1998, \mnras, 294, 139

\vfill\eject

\centerline {\bf FIGURE CAPTIONS}
\vskip1cm

\figcaption{Expected halo WDLF for pure H-atmosphere WD's in the V-band, for three different
halo ages and for $X_{WD}$=50\%. Solid line : IMF1; dashed line : IMF2 for $t_h=14$ and 16 Gyr.
The solid circles correspond to the faint end of the
disk WDLF (Liebert et al., 1998) while
the squares
correspond to the WD's in the sample with
$v_{tan}>250\,\kms$}
\vskip1cm

\figcaption{M$_V$ vs (V-I) color-magnitude diagram
for three different WD masses scaning the standard C/O
WD mass range with ages in Gyr indicated by the filled circles for the mean value 0.8 $\msol$}
\vskip1cm

\figcaption{Discovery functions in the V-band for 2 magnitude limits, namely
$m_V=22$ (top lines) and $m_V=20$ (bottom lines), and for 2 halo ages,
for $X_{WD}$=50\%.
Solid and long-dash lines : IMF1; dot and short-dash lines : IMF2.
Note that a kinematically selected
sample might reduce the explored volume and thus the density of
observable
WD's as a function of the magnitude.}

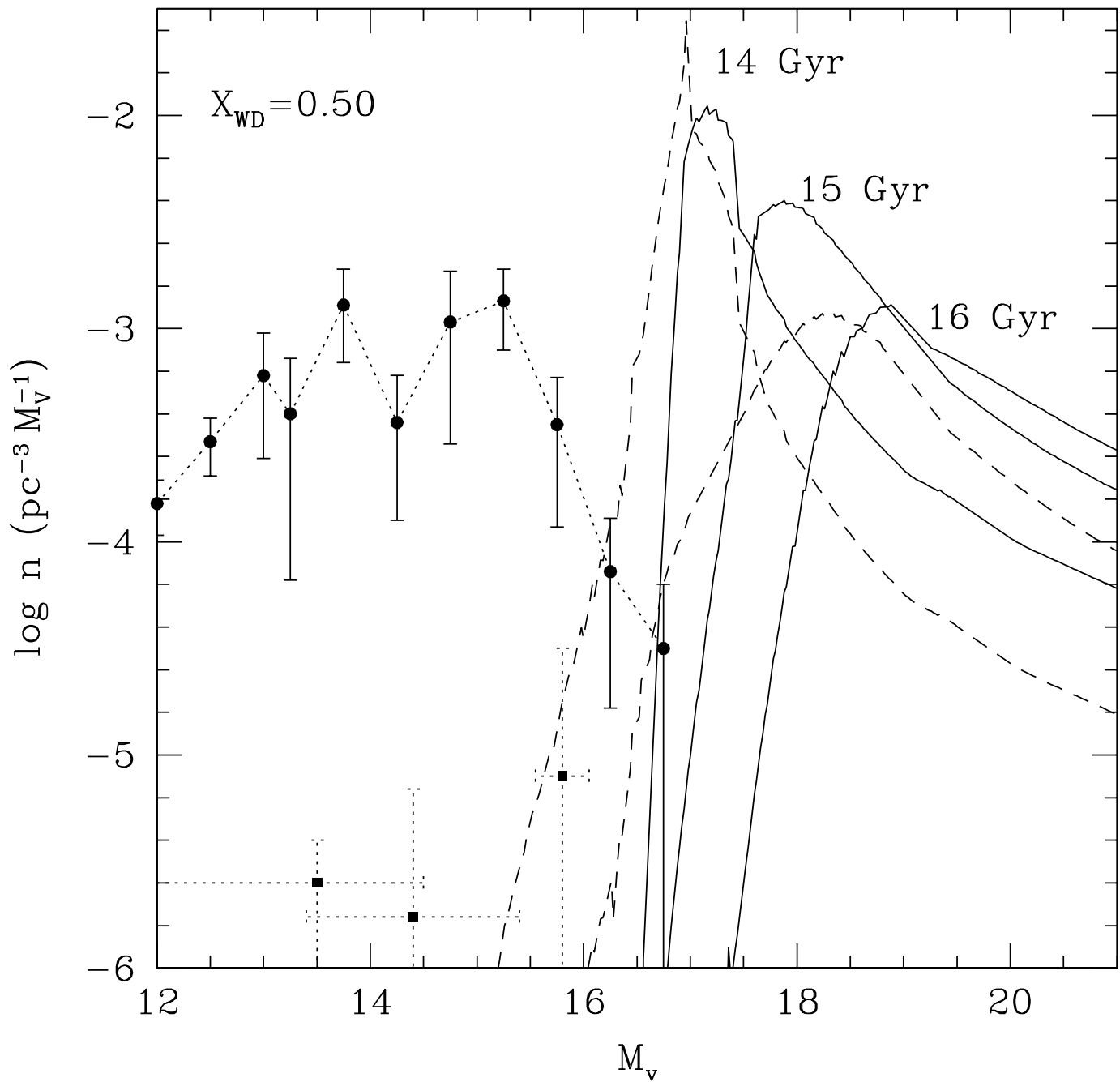
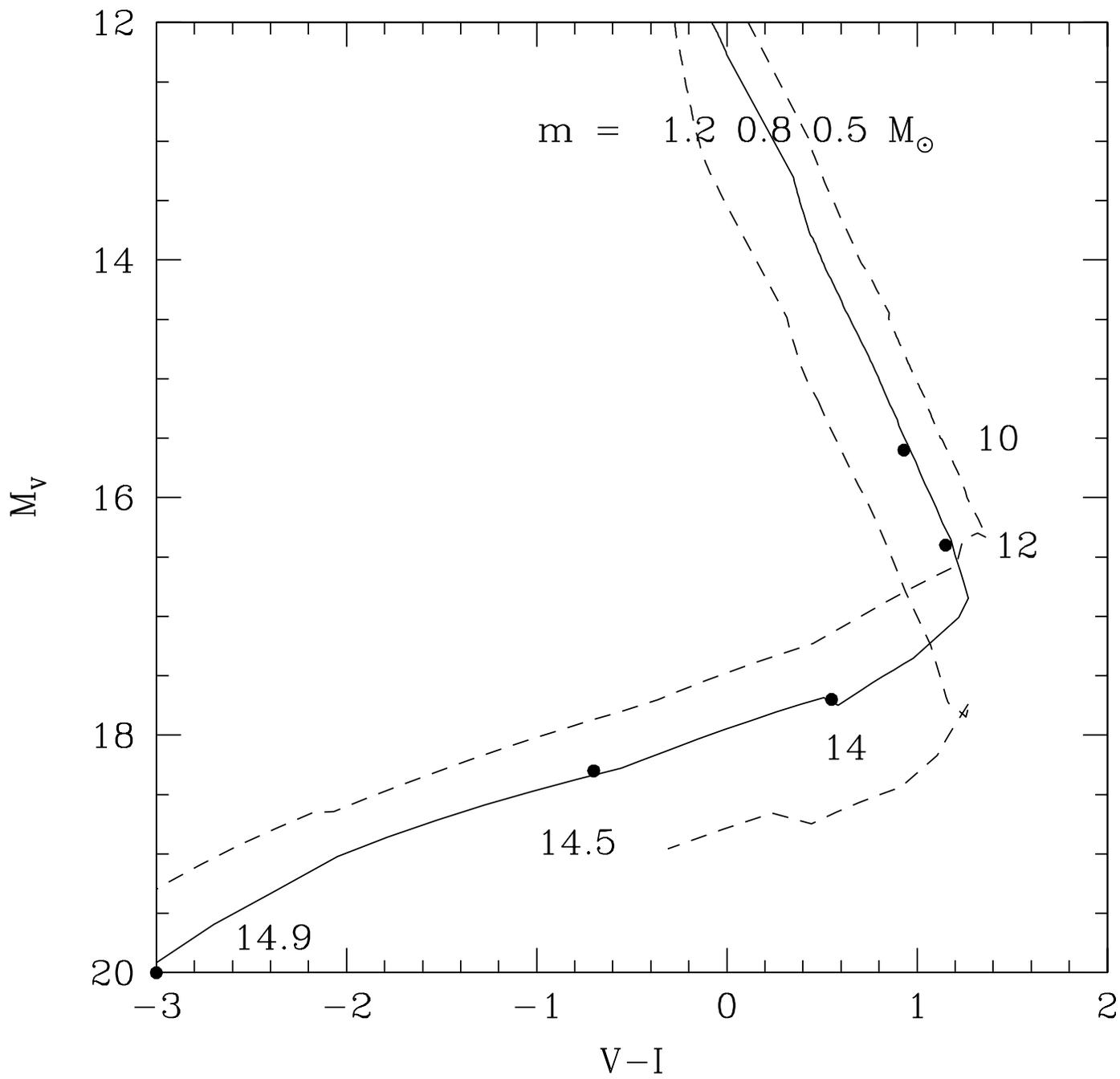
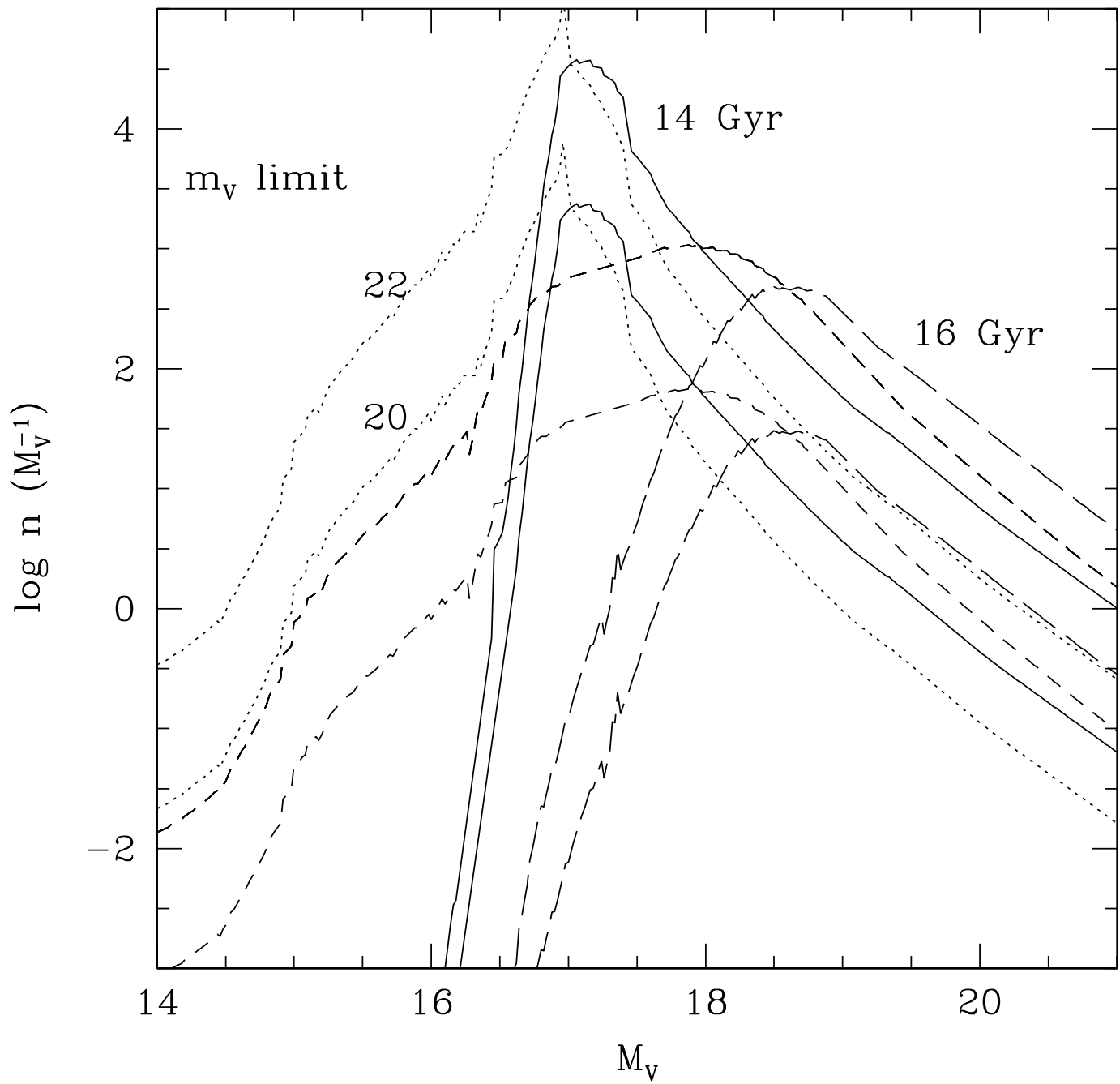
\end{document}